\documentclass[aps,pra,twocolumn,showpacs,superscriptaddress,nofootinbib,longbibiliography,10pt]{revtex4-2}
\usepackage{graphicx,amsmath,amsfonts,amssymb,braket,bbm}
\usepackage[dvipsnames]{xcolor}

\definecolor{ibmblue}{HTML}{0F62FE}

\usepackage{printlen}
\usepackage[colorlinks,linkcolor=ibmblue,citecolor=ibmblue,urlcolor=ibmblue]{hyperref}
\usepackage[capitalise]{cleveref}

\newcommand{\crt}[1]{\hat{a}^\dagger_{#1}}
\newcommand{\dst}[1]{\hat{a}^{\phantom{\dagger}}_{#1}}
\newcommand{\bts}[1]{{\bf{#1}}}
\newcommand{\numberop}[1]{\hat{n}_{#1}}
\newcommand{\device}[1]{$\mathsf{ibm\_#1}$}
\newcommand{\scientific}[2]{$#1 \cdot 10^{#2}$}

\newcommand{\revision}[1]{\textcolor{black}{#1}}

\date{\today}

\begin{document}

\title{Quantum-centric computation of molecular excited states \\ with extended sample-based quantum diagonalization}

\author{Stefano Barison}
\email{stefano.barison@epfl.ch}
\affiliation{Institute of Physics, \'{E}cole Polytechnique F\'{e}d\'{e}rale de Lausanne (EPFL), CH-1015 Lausanne, Switzerland}
\affiliation{Center for Quantum Science and Engineering, EPFL, Lausanne, Switzerland}
\affiliation{National Centre for Computational Design and Discovery of Novel Materials MARVEL, EPFL, Lausanne, Switzerland}

\author{Javier Robledo Moreno}
\email{j.robledomoreno@ibm.com}
\affiliation{IBM Quantum, IBM T. J. Watson Research Center, Yorktown Heights, NY 10598, USA}

\author{Mario Motta}
\email{mario.motta@ibm.com}
\affiliation{IBM Quantum, IBM T. J. Watson Research Center, Yorktown Heights, NY 10598, USA}

\begin{abstract}
The simulation of molecular electronic structure is an important application of quantum devices.
Recently, it has been shown that quantum devices can be effectively combined with classical supercomputing centers in the context of the sample-based quantum diagonalization (SQD) algorithm.
This allowed the largest electronic structure quantum simulation to date (77 qubits) and opened near-term devices to practical use cases in chemistry toward the hundred-qubit mark.
However, the description of many important physical and chemical properties of those systems, such as photo-absorption/-emission, requires a treatment that goes beyond the ground state alone.
In this work, we extend the SQD algorithm to determine low-lying molecular excited states.
The extended-SQD method improves over the original SQD method in accuracy, at the cost of an additional computational step.
It also improves over quantum subspace expansion based on single and double electronic excitations, a widespread approach to excited states on pre-fault-tolerant quantum devices, in both accuracy and efficiency.
We employ the extended SQD method to compute the first singlet (S$_1$) and triplet (T$_1$) excited states of the nitrogen molecule with a correlation-consistent basis set, and the ground- and excited-state properties of the [2Fe-2S] cluster. 
\end{abstract}

\maketitle

\section{Introduction}

The properties of matter are the results of interactions among its fundamental constituents, namely atoms and molecules.
Their behavior is governed by quantum mechanics and is connected to the electronic structure problem, which involves solving the time-independent Schr\"{o}dinger equation.
Determining its lowest-energy solution (i.e. the ground state of electrons at a given nuclear geometry) provides important understandings of the equilibrium properties of chemical compounds, including chemical reactivity, product distributions, and reaction rates. 

However, the study of the ground state alone is not exhaustive. Many processes -- such as the photochemical phenomena involving absorption/emission of photons and transitions between different many-electron wavefunctions -- require the inclusion of excited states to be described correctly \cite{Anderson1999,Polivka2004,Cheng2009,Johnson2015}.
Moreover, excited states usually possess symmetries that are different from the ground state ones, and may undergo chemical reactions that would be otherwise forbidden \cite{Woodward1965,Seder1986}.

Modeling these phenomena  by solving exactly the electronic Schr\"{o}dinger equation presents a considerable challenge even for contemporary (super)computers, currently limited to system sizes of 26 electrons in 23 orbitals~\cite{gao2024distributed}. 
As a result, it becomes necessary to employ approximate techniques, each with a specific domain of applicability, balancing between accuracy and computational cost~\cite{runge1984density,reining2002excitonic,sham1966many,hanke1980many,onida2002electronic,emrich1981extension,stanton1993equation,mcclain2016spectral,mcclain2017gaussian,ma2013excited,blunt2015excited,roos1980complete1,roos1980complete2,schmidt1998construction}.

Quantum computers have emerged as an alternative and complementary platform to simulate ground- and excited-state properties of chemical systems. 
The linear scaling between number of qubits and number of orbitals, combined with the polynomial cost of simulating the time-dependent Schr\"{o}dinger equation, offers new algorithms (e.g. phase estimation, adiabatic state preparation, and quantum filter diagonalization) to simulate electronic systems \cite{AspuruGuzik_2005, childs2018,farhi2000adiabatic,Babbush2014,Tameem2018,Kirby2023exactefficient,motta2024subspace}. 
While these algorithms have favorable asymptotic scaling, their performance and cost are also affected by constant prefactors, determining the size and depth of the corresponding quantum circuits as well as their accuracy.

As a result, experimental demonstration on near-term quantum processing units (QPUs) have focused on heuristic methods, e.g. based on the variational principle~\cite{Peruzzo_2014,rubin2016hybrid,Motta_2019,Eddins_2022,Zhang2022_prl}, that may deliver accurate results in certain regimes and account for hardware compatibility in their design. 
In this context, usually a QPU executes a subroutine in a larger program managed by a classical computer. 
Although these approaches have been applied to the simulation of chemical and physical systems, calculations have remained confined to $\sim$2-16 orbitals.

These limitations originate in part from the substantial overhead of quantum measurements affecting QPUs~\cite{wecker2015progress,gonthier2022measurements,Mazzola2024} and to the sensitivity of these measurements to errors, which prompted to reconsider the interaction between classical and quantum devices through the quantum-centric (super)computing model~\cite{alexeev2024quantum}. Pre-, peri- and post--processing by classical computers allows to execute a limited number of large quantum circuits on QPUs, and to mitigate errors at the level of individual samples from quantum measurements, giving access to large and more challenging instances of the electronic structure problem.
For example, the Sample-based Quantum Diagonalization algorithm (SQD)~\cite{kanno2023qsci,robledomoreno2024, kaliakin2024supramolecular} uses a QPU to sample electronic configurations and a classical computer to recover signal from such noisy data and to diagonalize the Hamiltonian in a subspace defined by the recovered configurations~\cite{kanno2023qsci}. SQD allowed ground-state simulations in active spaces of up to 36 orbitals, using up to 77 qubits and 10570 quantum gates~\cite{robledomoreno2024}. However, its application to electronic excited states is yet to be established.

Here, we present an extension of the SQD algorithm (Ext-SQD) to compute excited states of electronic systems.
First, we compare the method to other approaches, assessing its accuracy and efficiency.
Then, we use Ext-SQD to simulate the ground and low-lying excited states of N$_2$ along breaking of the triple bond in a correlation-consistent cc-pVDZ basis set.
The experiment is performed using 58 qubits on a Heron quantum processor
and 5204 (1792 2-qubit) quantum gates.
Finally, we apply the Extended-SQD to the study of the active space electronic structure of a methyl-capped [2Fe-2S] cluster \cite{sharma2014low,li2017spin} using a (30e,20o) active space, and was simulated using 45 qubits on a Heron quantum processor and 3170 quantum gates (of which 1100 are 2-qubit gates) .
The structure of this work is as follows: In \cref{sec:methods} we introduce different approaches to excited states determination and present our extension to the SQD algorithm.
Then, in \cref{sec:results} we present the results of our classical simulations and QPU experiments.
Finally, we conclude in \cref{sec:conclusion} with our considerations and outlooks on the proposed algorithm.

\section{Methods}
\label{sec:methods}

We start from the Born-Oppenheimer Hamiltonian in second quantization,
\begin{equation}
\label{eq:hamiltonian}
\hat{H} = E_0 + \sum_{\substack{pr \\ \sigma}} h_{pr} \crt{p\sigma} \dst{r\sigma} + \sum_{\substack{prqs\\\sigma\tau}} \frac{(pr|qs)}{2} \crt{p\sigma} \crt{q\tau} \dst{s\tau} \dst{r\sigma}
\;,
\end{equation}
where $p,r,q,s$ label spatial orbitals $\varphi_p$ in a finite orthonormal basis of $M$ elements,
$\sigma,\tau \in \{\alpha,\beta\}$  label spin polarizations, $E_0$ is an energy offset, and $h_{pr}$ and $(pr|qs)$ are the one- and two-electron integrals over the orbitals $\varphi_p$. Hartree units are used throughout, unless otherwise specified, and the numbers of spin- up and spin-down electrons are $N_\alpha$ and $N_\beta$, respectively. The goal of this study is to solve the time-independent Schr\"{o}dinger equation for the Hamiltonian Eq.~\eqref{eq:hamiltonian}, $\hat{H} \ket{\Phi_\mu} = \varepsilon_\mu \ket{\Phi_\mu}$, to determine its ground and low-lying excited states.

Here, we map the fermionic degrees of freedom of Eq.~\eqref{eq:hamiltonian} to qubits with a Jordan-Wigner (JW) transformation \cite{ortiz2002simulating,somma2002simulating,somma2005quantum}. In the JW mapping, the single-qubit basis states $\ket{0}/\ket{1}$ represent empty/occupied spin-orbitals, and therefore a computational basis state, i.e. $\ket{\bts{x}}$ with $\bts{x} \in \{0,1\}^{2M}$, represents a Slater determinant of the form
\begin{equation}
\label{eq:slater_bts}
\ket{\bts{x}} = \prod_{p\sigma} \left(\crt{p\sigma}\right)^{x_{p\sigma}} \ket{\varnothing}
\;,
\end{equation}
where $\ket{\varnothing}$ is the vacuum state. The wavefunction Eq.~\eqref{eq:slater_bts} has $N_\sigma$ spin-$\sigma$ electrons if $N_\sigma(\bts{x}) = \sum_p x_{p\sigma} = N_\sigma$, i.e., the first and second half of the bitstring $\bts{x}$ have Hamming weights $N_\alpha$ and $N_\beta$, respectively. Therefore, the target Hamiltonian eigenstates are linear combinations $\ket{\Phi_\mu} = \sum_{\bts{x}} c_{\bts{x}\mu} \ket{\bts{x}}$ of computational basis states with such properties. In this work, we will call the states in Eq.~\eqref{eq:slater_bts} ``configurations'', although in general they are not configuration state functions, i.e., eigenstates of the total spin operator.

\subsection{Excited-state methods}
\label{subsec:exc-state methods}

\textit{SQD -- }
The first method used in this work to study Hamiltonian excited states is the Sample-based Quantum Diagonalization (SQD)~\cite{robledomoreno2024,kanno2023quantum} itself.
Within SQD, a quantum circuit $\Psi$ is used to sample a set $\tilde{\mathcal{X}} = \{ \bts{x}^\prime_1 \dots \bts{x}^\prime_N \}$ of computational basis states.
On a pre-fault-tolerant device, even if the the quantum circuit $\Psi$ does not break the particle-number and spin-$z$ symmetries of Eq.~\eqref{eq:hamiltonian}, due to decoherence, $\tilde{\mathcal{X}}$ may contain configurations with incorrect particle numbers, $N_\sigma(\bts{x}^\prime_k) \neq N_\sigma$ for some $\bts{x}_k \in \tilde{\mathcal{X}}$. 
To overcome this limitation, Ref.~\cite{robledomoreno2024} proposed a self-consistent configuration recovery (S-CORE), that allows to transform $\tilde{\mathcal{X}}$ into a self-consistently determined set of configurations $\mathcal{X}_\mathrm{R} = \{ \bts{x}_1 \dots \bts{x}_N \}$ with correct particle number, i.e. $N_\sigma(\bts{x}_k) = N_\sigma$ for all $\bts{x}_k \in \mathcal{X}_\mathrm{R}$.

To approximate the ground-state wave function, $K$ sets of configurations or ``batches'' are randomly sampled from $\mathcal{X}_\mathrm{R}$, and each batch is used to produce a set $S^{(b)} = \{ \bts{y}^{(b)}_1 \dots \bts{y}^{(b)}_D \}$ of configurations closed under spin inversion symmetry~\cite{robledomoreno2024}. Note that, for this reason, $D$ can be substantially larger than $|\tilde{\mathcal{X}}|$. Within each batch, the ground-state wave function is approximated as
\begin{equation}
\label{eq:sqd_ground_state}
\ket{\tilde{\Phi}^{(b)}_0} = \sum_{k=1}^D c^{(b)}_{k0} \ket{\bts{y}^{(b)}_k}
\;,
\end{equation}
where the coefficients $c^{(b)}_{k0}$ are determined by solving the eigenvalue equation
\begin{equation}
\label{eq:sqd_coefficients}
\sum_{k=1}^D H^{(b)}_{kl} c^{(b)}_{k0} = \tilde{\varepsilon}^{(b)}_0 c^{(b)}_{l0} 
\;,\;
H^{(b)}_{kl} = \bra{\bts{y}^{(b)}_l}\hat{H}\ket{\bts{y}^{(b)}_k} 
\;,\;
\end{equation}
and $\tilde{\varepsilon}^{(b)}_0$ is the lowest eigenvalue of $H^{(b)}$. The wave function Eq.~\eqref{eq:sqd_ground_state} leading to the lowest energy $\tilde{\varepsilon}^{(b)}_0$ is chosen as ground-state approximation. We will henceforth denote the corresponding set of configurations, eigenvalue, and eigenvector as $S = \{ \bts{y}_1 \dots \bts{y}_D \}$, $\ket{\tilde{\Phi}_0}$, and $\tilde{\varepsilon}_0$ respectively.

A simple and relatively economical way to approximate low-lying excited states of Eq.~\eqref{eq:hamiltonian} within SQD is to compute multiple eigenpairs of the matrix $H$, which yields wave functions of the form
\begin{equation}
\label{eq:sqd_excited_states}
\ket{\tilde{\Phi}_\mu} = \sum_{k=1}^D c_{k\mu} \ket{\bts{y}_k}
\;.
\end{equation}
This procedure, though natural and compelling, has an important limitation: configurations in $\tilde{\mathcal{X}}$ are drawn from a quantum circuit $\Psi$ that approximates the ground state of $\hat{H}$, and configurations $\mathcal{X}_\mathrm{R}$ are derived from $\tilde{\mathcal{X}}$ through a S-CORE procedure guided by approximate ground-state wavefunctions. These properties may strongly bias SQD towards the ground-state wave function leading, e.g., to overestimate excitation energies. Furthermore, when a Hamiltonian has a molecular point-group symmetry (e.g. $\mathrm{C_{2v}}$ or $\mathrm{D_{\infty h}}$), the set $\mathcal{X}_\mathrm{R}$ contains configurations in a specific irrep of the symmetry group
(e.g. $\mathrm{A_1}$ or $\mathrm{A_{1g}}$ respectively) and excited states in different irreps are not accessible within SQD.

\textit{QSE(SD) -- }An alternative to SQD is a quantum subspace expansion (QSE) \cite{mcclean2017hybrid,motta2024subspace} based on single and double electronic excitations. In this framework, Hamiltonian eigenstates are approximated by states of the form
\begin{equation}
\label{eq:qse_eigenstates}
\ket{\tilde{\Phi}_\mu} = \sum_I d_{I\mu} \, \hat{E}_I \ket{\tilde{\Phi}_0}
\end{equation}
where $\hat{E}_I \in \{ \mathbbm{I} , \crt{a\sigma} \dst{i\sigma} , \crt{a\sigma} \crt{b\tau} \dst{j\tau} \dst{i\sigma} \}$ is a set of excitation operators -- here, single and double excitations from occupied to virtual orbitals, respectively $ij$ and $ab$, defined by the restricted closed-shell Hartree-Fock state -- and $\ket{\tilde{\Phi}_0}$ is the SQD ground-state approximation Eq.~\eqref{eq:sqd_ground_state}.
The coefficients $d_{I\mu}$ are determined variationally, solving the generalized eigenvalue equation $M d = S d \tilde{\varepsilon}$ with
\begin{equation}
\label{eq:qse_eigh}
M_{IJ} = \bra{\tilde{\Phi}_0} \hat{E}_I^\dagger \hat{H} \hat{E}_J \ket{\tilde{\Phi}_0}
\;,\;
S_{IJ} = \bra{\tilde{\Phi}_0} \hat{E}_I^\dagger \hat{E}_J \ket{\tilde{\Phi}_0}
\;.
\end{equation}
This method, here abbreviated as QSE(SD), has been used to simulate molecular excited states on quantum hardware in the last few years \cite{colless2018computation,Huang2022simulating,Khan2022sim,dhawan2023quantum,tammaro2023n,huang2023quantum,motta2023quantum,castellanos2023quantum}, in part because it does not require running deeper circuits than those needed to prepare $\ket{\tilde{\Phi}_0}$. However, QSE(SD) has a substantial measurement overhead, as it requires estimating high-order $k$-body reduced density matrices. Furthermore, it shares the accuracy limitations of truncated configuration interaction methods, e.g. the lack of size-extensitivity and size-consistency of ground-state energies and lack of size-intensivity of excitation energies \cite{motta2024subspace}.

\textit{Ext-SQD -- }
A third path toward to compute molecular excited states is an extension of the SQD method, rather naturally suggested by the QSE(SD) approach. The approximate eigenfunctions in Eq.~\eqref{eq:qse_eigenstates} have the following form, readily obtained from Eq.~\eqref{eq:sqd_ground_state},
\begin{equation}
\label{eq:toward_ext_sqd_1}
\ket{\tilde{\Phi}_\mu} = \sum_{Ik} d_{I\mu} c_{k0} \, \hat{E}_I \ket{\bts{y}_k}
\;.
\end{equation}
Since the action of an excitation operator $\hat{E}_I$ on a computational basis state $\ket{\bts{y}_k}$ yields another computational basis state,
\begin{equation}
\label{eq:extension_of_configurations}
\hat{E}_I \ket{\bts{y}_k} = \gamma_{Ik} \ket{\bts{z}_{Ik}}
\;,\;
\gamma_{Ik} \in \{0,1,-1\}
\;,
\end{equation}
equation~\eqref{eq:toward_ext_sqd_1} takes the form
\begin{equation}
\label{eq:toward_ext_sqd_2}
\ket{\tilde{\Phi}_\mu} = \sum_{Ik} d_{I\mu} \gamma_{Ik} c_{k0} \, \ket{\bts{z}_{Ik}}
\;.
\end{equation}
Equation~\eqref{eq:toward_ext_sqd_2} shows that $\ket{\tilde{\Phi}_\mu}$ is a linear combination of computational basis state in the extended set $S_\mathrm{E} = \{ \bts{z}_{Ik} \}_{Ik}$, with particularly structured expansion coefficients $d_{I\mu} \gamma_{Ik} c_{k0}$. In the light of Eq.~\eqref{eq:toward_ext_sqd_2}, it is natural to consider approximations to excited states of the form
\begin{equation}
\label{eq:toward_ext_sqd_3}
\ket{\tilde{\Phi}_\mu} = \sum_{\bts{z} \in S_\mathrm{E}} f_{\bts{z} \mu} \ket{\bts{z}}
\;,
\end{equation}
i.e., linear combinations of computational basis state in $S_\mathrm{E}$ with coefficients determined variationally exactly as in SQD, Eq.~\eqref{eq:sqd_coefficients}. We call this approach Extended SQD (Ext-SQD), and employ it along SQD and QSE(SD) in the reminder of this work. 

Compared to SQD, Ext-SQD is by construction more accurate and expensive, because it uses configurations in $S_\mathrm{E}$ rather than $\mathcal{X}_\mathrm{R}$. The increase in computational cost is relatively modest, since $D_E = |S_\mathrm{E}|$ is bounded by $D \prod_\sigma N_\sigma (M-N_\sigma)$.
Compared to QSE(SD), Ext-SQD is by construction more accurate because, while both use computational basis states in $S_\mathrm{E}$, Ext-SQD has expansion coefficients, Eq.~\eqref{eq:toward_ext_sqd_3}, free from the particular form in Eq.~\eqref{eq:toward_ext_sqd_2}, which is a source of greater variational flexibility. Furthermore, Ext-SQD is less expensive than QSE(SD), because it does not require the computation of high-order $k$-body reduced density matrices.
It is also less prone to numeric instability, because it requires solving a standard eigenvalue equation rather than a generalized eigenvalue equation, Eq.~\eqref{eq:qse_eigh}, where the overlap matrix in Eq.~\eqref{eq:qse_eigh} may be ill-conditioned.

\subsection{Computational details}
\label{subsec:compl details}

\subsubsection{Active spaces}
\label{sec:details_1}

For the dissociation of N$_2$, we used the classical electronic structure PySCF~\cite{sun2018pyscf,sun2020recent} to generate optimized mean-field orbitals (of restricted closed-shell Hartree-Fock, RHF, type) and matrix elements of the Hamiltonian Eq.~\eqref{eq:hamiltonian} in: 
\begin{enumerate}
\item a (10e,8o) active space obtained from the cc-pVDZ basis set~\cite{dunning1989gaussian} using the automated valence active-space (AVAS) procedure~\cite{sayfutyarova2017automated}; we selected active-space orbitals overlapping with nitrogen 2s and 2p atomic orbitals
\item a (10e,16o) active space obtained from the 6-31G basis set~\cite{hehre1972self} using the frozen-core approximation (results from this active space are presented exclusively in the Appendix)
\item a (10e,26o) active space obtained from the cc-pVDZ basis set using the frozen-core approximation
\end{enumerate}

For the [2Fe-2S] cluster, we used a publicly available (30e,20o) active space~\cite{FeSRepo} comprising iron 3d, sulfur 3p, and cluster-ligand bonding orbitals~\cite{sharma2014low,li2017spin}.

The SQD and Ext-SQD calculations are carried out using the Qiskit-addon-sqd~\cite{sqd_addon} package.
To provide a reference against which to interpret SQD, QSE(SD), and Ext-SQD results, we computed ground- and excited-state energies with the Complete Active-Space Configuration Interaction (CASCI) and Heat-Bath Configuration Interaction (HCI) \cite{holmes2016heat} as implemented in PySCF.

\subsubsection{Quantum circuits}

Having selected a set of single-electron orbitals for each of the studied species, we constructed quantum circuits from which to measure configurations. Following~\cite{robledomoreno2024}, in this work we employed a truncated version of the local unitary cluster Jastrow (LUCJ) ansatz~\cite{motta2023bridging},
\begin{equation}
\label{eq:lucj}
\ket{\Psi} = e^{-\hat{K}_1} e^{\hat{K}_0} e^{i \hat{J}_0} e^{-\hat{K}_0} \ket{\bts{x}_{\mathrm{RHF}}}
\;,
\end{equation}
where $\bts{x}_{\mathrm{RHF}}$ is the bitstring representing the RHF state in the JW mapping and
\begin{equation}
\hat{K}_\mu = \sum_{\substack{pr \\ \sigma}} K_{pr}^\mu \, \crt{p \sigma} \dst{r \sigma}
\;,\;
\hat{J}_\mu = \sum_{\substack{pr \\ \sigma\tau}} J_{p\sigma, r\tau}^\mu \, \numberop{p \sigma} \numberop{r \tau}
\;,
\end{equation}
are a generic one-body operator and a density-density operator, respectively. The locality approximation underlying LUCJ involves restricting the density-density operators, i.e., sparsifying the matrices $J^\mu$
to include only spin-orbitals mapped to qubits that are adjacent in the topology of a given device~\cite{motta2023bridging}. Alternatively, one could include spin-orbitals mapped to qubits that can be coupled using SWAP networks compatible with the device's topology.
Following~\cite{robledomoreno2024} we performed optimization-free experiments, leveraging the connection between LUCJ and classical coupled-cluster theory (CCSD) to parametrize the circuits in Eq.~\eqref{eq:lucj} using $t_2$ amplitudes  from a CCSD calculation, however performing a quantum-classical optimization could improve the quality of the results.

The LUCJ circuits were generated using the \texttt{ffsim} library \cite{ffsim} and executed on IBM's 133-qubit Heron superconducting quantum processor \device{torino}, using twirled readout error mitigation (ROEM)~\cite{nation2021scalable} and dynamical decoupling (DD)~\cite{viola1998dynamical,kofman2001universal,biercuk2009optimized,niu2022effects} to mitigate errors arising from qubit measurements and quantum gates, respectively. 
We employed the implementation of ROEM and DD available through the \texttt{SamplerV2} primitive of Qiskit's~\cite{Qiskit} \texttt{Runtime} library. See \cref{app:exp_details} for additional details.

\subsubsection{Details of SQD, QSE(SD), and Ext-SQD calculations}
\label{sec:details_3}

After executing the quantum circuit Eq.~\eqref{eq:lucj}, we run SQD to approximate the molecular ground state. The number of configurations gathered, iterations of S-CORE, number of batches, dimension of the SQD subspace, Ext-SQD, and number of single and double excitations (specifying the cost of QSE(SD)) for N$_2$ are listed in Table~\ref{tab:sqd}.

\begin{table}[ht!]
\begin{tabular}{ccccccc}
\hline\hline
system & $|\tilde{\mathcal{X}}|$ & S-CORE & $K$ & $D$ & $D_E$ & $(n_s,n_d)$ \\
\hline
N$_2$ (10e,8o)  & $10^3$ &  5 & 10 & $2.5 \cdot 10^3$ & $3 \cdot 10^3$   & (56,210) \\
N$_2$ (10e,16o) & $10^5$ & 10 & 10 & $4 \cdot 10^6$ & $2.1 \cdot 10^6$ & (272,5460) \\
N$_2$ (10e,26o) & $10^5$ & 10 & 10 & $9 \cdot 10^6$ & $4.7 \cdot 10^6$ & (702,44850) \\
\hline\hline
\end{tabular}
\caption{ 
    \textbf{Summary of the N$_2$ Ext-SQD experiments.}
    The table shows the different active spaces employed in the SQD calculations of N$_2$.
    $|\tilde{\mathcal{X}}|$ indicates the amount of samples (either classical or quantum) collected, while S-CORE and $K$ the number of configuration recovery iterations run and the number of different batches used, respectively.
    $D$ indicates the final dimension of the subspace in which the classical solver is run.
    In case of multiple instances, the maximum $D$ is shown.
    Finally,  $(n_s,n_d)$ indicates the number of single and double excitations operators used in the excited-state calculations, while $D_E$ the corresponding extended subspace dimension.
    While extended, $D_E$ might be smaller than $D$ in some instances (see \cref{app:tech_details} for additional information).
 }
\label{tab:sqd}
\end{table}

We use the classical solver to determine not only the ground state, but also the eigenpairs of the first two excited states, $T_1$ and $S_1$.
$T_1$ is the first triplet state, while $S_1$ is the first singlet excited state.
QSE(SD) is run with all single and double excitation operators, forming square matrices $M_{IJ}$ and $S_{IJ}$ of \cref{eq:qse_eigh} of dimension $(n_s+n_d+1) \times (n_s+n_d+1)$.
Finally, Ext-SQD is performed by applying all single and double excitation operators to the SQD wave function.

 For the [2Fe-2S] cluster, $|\tilde{\mathcal{X}}| = 2.4576 \cdot 10^6$, $\textrm{S-CORE} = 7$, $D = 16.7 \cdot 10^6$, $D_E=126.2 \cdot 10 ^6$ and $(n_s, n_d, n_t) = (420, 14535, 610470)$, where $n_t$ is the number of three-particle excitations.

\section{Results}
\label{sec:results}

\subsection{Assessment of accuracy}

\begin{figure}[ht]
    \includegraphics[width=1\columnwidth]{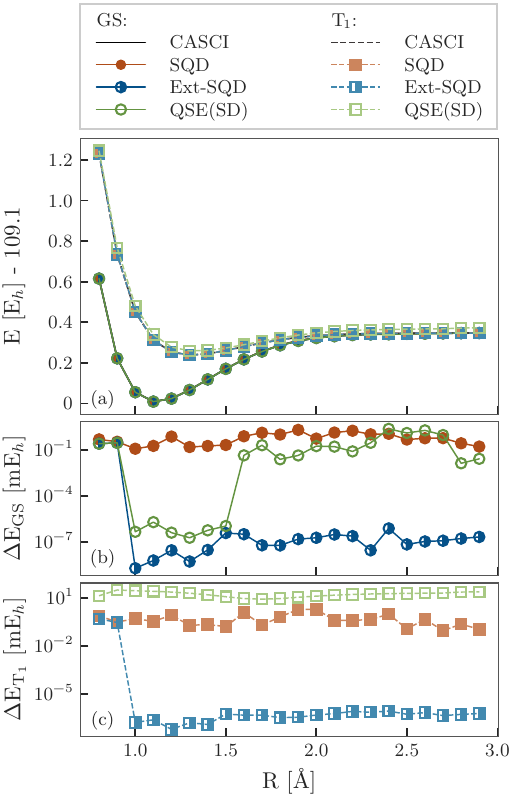} 
      \caption{
      \textbf{Ext-SQD assessment of accuracy.}
      \textbf{(a)} Ground-state (GS) and first excited-state (T$_1$) potential energy surfaces of N$_2$ in a (10e,8o) active space using the SQD method (SQD), its extension (Ext-SQD), the quantum subspace expansion with single and double excitation operators (QSE(SD)), and classical diagonalization (CASCI). 
      \textbf{(b,c)} Deviations $\Delta E$ from CASCI in the total energy, for SQD, Ext-SQD, and QSE(SD), along the GS (b) and T$_1$ (c) potential energy curves.
      }
      \label{fig:diss_N2_qse}
\end{figure}

As highlighted in \cref{subsec:exc-state methods}, there are several approaches to compute excited states based on SQD. We compare them in Figure~\ref{fig:diss_N2_qse}, focusing on the dissociation of N$_2$ molecule in the (10e,8o) active space described in Sec.~\ref{sec:details_1}.
The calculations in Figure~\ref{fig:diss_N2_qse} use a classical device to draw configurations from the uniform probability distribution, and perform SQD calculations as detailed in Sec.~\ref{sec:details_3}.
We use QSE(SD), Ext-SQD, and CASCI to determine the ground state (GS) and the first excited state (T$_1$) of the molecule. The latter is the lowest-energy triplet state.
As highlighted in the middle and bottom panels, both QSE(SD) and Ext-SQD deviate from FCI by less than 1 mHa across dissociation. On the other hand, QSE(SD) overestimates the energy of T$_1$ by 10-30 mHa whereas Ext-SQD agrees with CASCI within 1 mHa.

While QSE(SD) requires solving an eigenvalue equation of size 267, smaller than Ext-SQD, the computational cost of forming the Hamiltonian and overlap matrix is higher than that of producing the Ext-SQD wavefunction.
Indeed, every point of the dissociation curve for QSE(SD) required 5 hours running on 30 cores on a bare metal node consisting on four sockets each with a Intel Xeon Platinum 8260 (2.40GHz) processor, while the Ext-SQD required on average 20s per point on a single core of an M3 Pro laptop processor.
Furthermore, as $R$ increases beyond 2.5 Angstrom, the QSE(SD) overlap matrix is more likely to become ill-conditioned, requiring regularization of the eigenvalue equation. For this reason, QSE(SD) energies are above SQD in that regime. 
Due to the lower accuracy and higher computational cost of QSE(SD), we henceforth employ SQD and Ext-SQD.

\subsection{Applications: dissociation of N$_2$}

We now consider the dissociation of N$_2$ in a (10e,26o) active space. The spectrum of N$_2$ is complex, encompassing electronic transitions over a broad range of wavelengths. It is also thoroughly studied, both experimentally and in classical quantum chemistry \cite{Lofthus1977_JPCRD,Larsen2000_JCP,Ono2009_PSST,Gdanitz1998,Chan2002}, where is a well-known test of the accuracy of electronic structure methods in the presence of static electronic correlation~\cite{bulik2015can}.

\begin{figure*}[ht]
    \includegraphics[width=1\textwidth]{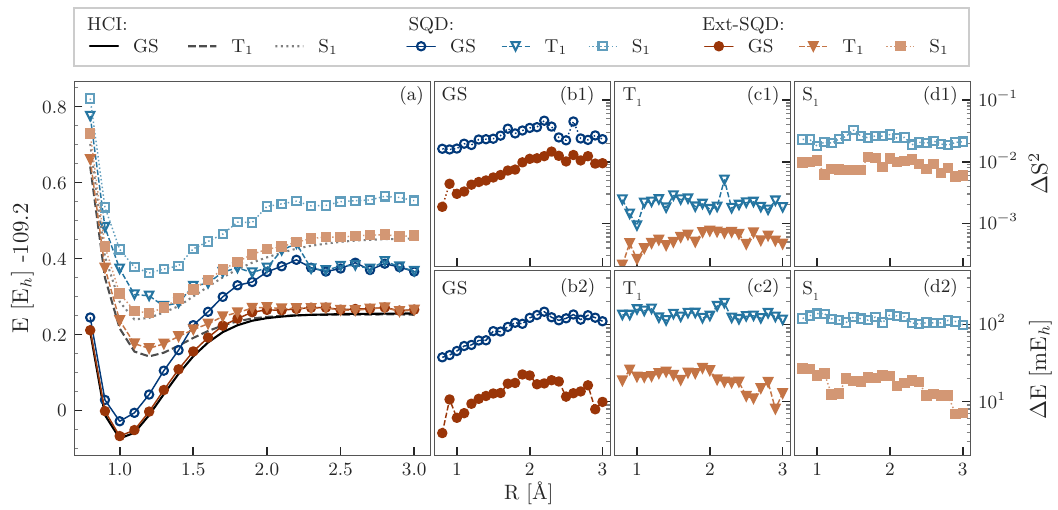} 
      \caption{
      \textbf{N$_2$ dissociation.}
      \textbf{(a)} Ground-state (GS) and first two excited-state potential energy curves (T$_1$ and S$_1$) of N$_2$ in the cc-pVDZ basis set from SQD, extended SQD (Ext-SQD), and heat-bath configuration interaction (HCI).
      \textbf{(b1, c1, d1)} Deviations $\Delta S^{2}$ from HCI in the total spin, for SQD and Ext-SQD, along the GS (b1), T$_1$ (c1), and $S_1$ (d1) potential energy curves.
      \textbf{(b2, c2, d2)} Deviations $\Delta E$ from HCI in the total energy, for for SQD and Ext-SQD, along the GS (b2), T$_1$ (c2), and $S_1$ (d2) potential energy curves.
      }
      \label{fig:diss_N2_ccpvdz}
\end{figure*}

The comparison between the SQD, Ext-SQD, and SCI is shown in Fig.~\ref{fig:diss_N2_ccpvdz}, focusing on the ground state, T$_1$, and S$_1$ excited states. As seen, Ext-SQD improves the quality of all three curves by order $10^2$ milliHartre, and also leads to less pronounced discontinuities as the bondlength $R$ varies. Ext-SQD excited states also feature values of $S^2$ much closer to the exact values (0 and 2 for singlet and triplet states, respectively) than SQD.

From the results in Fig.~\ref{fig:diss_N2_ccpvdz}, we can compute several quantities of interest for nitrogen, and compared it to classical simulations and experimental data. Table~\ref{tab:n2_quantities} reports the equilibrium bondlengths, vibrational frequencies, and dissociation energies of N$_2$ obtained with HCI, SQD and Ext-SQD.
We estimate equilibrium bondlengths and vibrational frequencies by fitting the potential energy curve in the interval $0.9 \, \mathrm{\AA} \leq R \leq 1.5 \, \mathrm{\AA}$ to a Morse potential.
We estimate dissociation energies by (1) fitting the potential energy curve to a power-law curve in the interval $R \leq 2.0 \, \mathrm{\AA}$, and (2) taking the difference between the asymptotic value of the energy (from the power-law fit) to the value of the energy at the equilibrium bondlength (from the Morse-potential fit).
Ext-SQD properties are in substantially better agreement with HCI than their SQD counterparts, illustrating how Ext-SQD can economically improve the SQD ground-state approximation without (or before) resorting to energy-variance extrapolations~\cite{robledomoreno2024}.

\begin{table*}[ht]
    \centering
    \begin{tabular}{l | c | c | c | c }
         \hline
         \hline
         \textbf{Quantity} & \textbf{Experiment} & \textbf{HCI}  & \textbf{SQD} & \textbf{Ext-SQD} \\
         \hline
         Equilibrium bondlength [Å]         &  1.097513         & 1.119256 & 1.109704 & 1.119951 \\
         Vibrational frequency  [cm$^{-1}$] &  2330.00          & 2319.24  & 2402.53  & 2319.01  \\
         Dissociation energy    [kJ/mol]    & 945.33 $\pm$ 0.59 & 865.50 $\pm$ 0.68   & 1041.56 $\pm$ 7.44 & 891.87 $\pm$ 3.93\\
         \hline
         \hline
    \end{tabular}
    \caption{ 
        \textbf{Properties from the N$_2$ potential energy curve.}
        Comparison of N$_2$ equilibrium bondlengths, vibrational frequency, and dissociation energy (rows 2-4) from experiment (column 2, from Refs.~\cite{irikura2007experimental,huber2013molecular}) and HCI, SQD, Ext-SQD (columns 3-5) in the cc-pVDZ basis set with frozen-core approximation. Statistical uncertainties in columns 3-5 reflect uncertainties from a standard fitting procedure.
        }
    \label{tab:n2_quantities}
\end{table*}

\subsection{Applications: the [2Fe-2S] cluster}

Given the complicated nature of the electronic-structure properties of the [2Fe-2S] cluster~\cite{sharma2014low}, we asses the accuracy of two flavors of Ext-SQD. The first is obtained by the action of all single- and double-particle transition operators (SD), as in Eq.~\eqref{eq:extension_of_configurations}, and the second by the action of all single-, double and triple-particle transition operators (SDT).

\begin{figure}[t!]
    \includegraphics[width=\columnwidth]{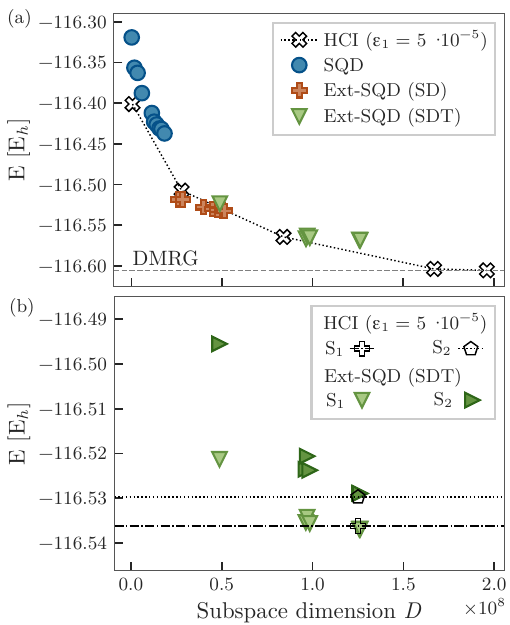} 
      \caption{
      \textbf{Ground- \revision{and excited-}state energies of the [2Fe-2S] cluster.}
      (a) Ground-state energy estimation as a function of the subspace dimension for SQD (blue circles), Ext-SQD with added single- and double-excitations (SD, orange pluses), and Ext-SQD with single-, double- and triple-excitations (SDT, green triangles). The energies and subspace dimensions are compared to those visited by different iterations (following the flow of the black dotted line) of HCI (black crosses). The truncation coefficient for the HCI calculation is indicated in the legend (see Ref.~\cite{Smith2017heat} for details). The gray horizontal line shows the DMRG energy reported in Ref.~\cite{li2017spin}.
      \revision{(b) Excited-state energies estimation as a function of the subspace dimension Ext-SQD with single-, double- and triple-excitations (SDT, downward triangles for $S_1$, rightward triangles for $S_2$). Energies and dimensions are compared to HCI results. Dotted and dash-dotted lines represent again the HCI results and are a guide to the eyes. The truncation coefficient for HCI is the same as in panel (a).}
      }
      \label{fig:2Fe2S}
\end{figure}

The comparison between HCI, SQD and Ext-SQD \revision{ground state energies as a function of the subspace dimension} is shown in \revision{panel (a) of} Fig.~\ref{fig:2Fe2S}. 
For a fixed subspace dimension, the ground-state energy estimations obtained from Ext-SQD (SD) and Ext-SQD (SDT) are in good agreement with those produced by HCI.
The passage from SQD to Ext-SQD (SD) decreases the deviation from the ground-state energy -- computed with density matrix renormalization group (DMRG)~\cite{li2017spin} -- from $\sim$$200$ to $\sim$$75$ milliHartree, while the number of configurations increases from $\sim$$15$ to $\sim$$35$ million. The inclusion of triple-particle transition operators further decreases the deviation to $\sim$$50$ milliHartree while increasing the number of configurations to $\sim$$100$ million.

\revision{Panel (b) of Fig.~\ref{fig:2Fe2S} compares the S$_1$ and S$_2$ energies and subspace dimensions obtained from Ext-SQD (SDT) to those obtained with HCI. The points of highest subspace dimension considered in the Ext-SQD (SDT) calculations show excellent agreement ($<1$ milliHartree) with the HCI estimations at comparable subspace dimensions. }

In the canonical interpretation of iron-sulfur clusters based on the Heisenberg-Double-Exchange (HDE) model, the Fe atoms are considered in definite valence states. For the oxidized [2Fe-2S]${}^{2-}$ cluster considered in this work, both Fe atoms are assumed high-spin Fe(III) with total spins $S_1,S_2 = 5/2$, and it is assumed that there is no double-exchange. Therefore, the HDE model reduces to the Heisenberg form $\hat{H} = 2J \vec{S}_1 \cdot \vec{S}_2$ where $J>0$ is a coupling constant. The ground state of the HDE Hamiltonian is low-spin, with maximal antiferromagnetic alignment.

In Table~\ref{tab:fe2s2_quantities}, we examine the ability of computational methods, including SQD and Ext-SQD, to describe the ground-state physical properties of the oxidized [2Fe-2S] cluster. Specifically, we compute the number of spin-$\sigma$ electrons on Fe atom $a=1,2$ as the expectation value of
\begin{equation}
\hat{N}^{(a)}_{\mathrm{e,\sigma}} =
\sum_{p \in \mathrm{Fe}_a} \numberop{p\sigma}
\;,\;
\end{equation}
where $p \in \mathrm{Fe}_a$ denotes the 3d orbitals of iron atom $a$, the local spin of Fe atom $a$ as the expectation value of
\begin{equation}
\hat{\bts{S}}^{(a)} = \frac{1}{2} \sum_{p \in \mathrm{Fe}_a} \bts{P}_{\sigma\tau} \, \crt{p\sigma} \dst{p\tau}
\;,\;
\end{equation}
where $\bts{P}$ is the vector of $2 \times 2$ Pauli matrices, and the spin-spin \revision{connected} correlation
\begin{equation}
C = \langle \Psi | \hat{\bts{S}}^{(1)} \cdot \hat{\bts{S}}^{(2)} | \Psi \rangle 
-
\langle \Psi | \hat{\bts{S}}^{(1)} | \Psi \rangle \cdot \langle \Psi |\hat{\bts{S}}^{(2)} | \Psi \rangle 
\;.
\end{equation}
Restricted HF, MP2, CCSD, and CISD predict $\sim$$6.2$ electrons on each Fe atom, zero local spins, and a very low antiferromagnetic ($C<0$) spin-spin correlation. Unrestricted methods predict $\sim$$5$ spin-up and $\sim$$1$ spin-down electrons on the left Fe atom and viceversa on the right Fe atom, local spins of roughly equal magnitude and opposite orientation (roughly $\pm 2$ for the left and right atom, respectively), and a very low antiferromagnetic spin-spin correlation ($C \simeq 0$ because $\langle \Psi | \hat{\bts{S}}^{(1)}  \cdot \hat{\bts{S}}^{(2)} | \Psi \rangle \simeq \langle \Psi | \hat{\bts{S}}^{(1)} | \Psi \rangle  \cdot \langle \Psi |\hat{\bts{S}}^{(2)} | \Psi \rangle $).
The behavior of restricted HF, MP2, CCSD, and CISD is a reflection of their inability to capture the strong electronic correlation arising from the composition of the local Fe spins into a global singlet state. Unrestricted methods exhibit spin symmetry breaking with non-zero expectation values for the local spin operators, whose opposite orientation is a reflection of the antiferromagnetic correlation between the two Fe atoms. The values of \revision{$\langle \Psi | \hat{\bts{S}}^{(1)} \cdot \hat{\bts{S}}^{(2)} | \Psi \rangle$} range between $-3.84$ and $-4.41$, in good agreement with the DMRG estimate of \revision{$\langle \Psi | \hat{\bts{S}}^{(1)} \cdot \hat{\bts{S}}^{(2)} | \Psi \rangle  = -4.92$}~\cite{sharma2014low}.
SQD and Ext-SQD predict a stronger antiferromagnetic correlation between the two Fe atoms than other restricted methods, with values of $-0.6315$ and $-1.7619$ respectively.
\revision{Ext-SQD prediction is in agreement with HCI result of $-1.7807$, which is obtained with a larger subspace dimension. Table~\ref{tab:fe2s2_exc_quantities} compares the charge and spin properties of the Fe atoms for the S$_1$ and S$_2$ excitations obtained from HCI and Ext-SQD (SDT). As it could be anticipated from the agreement in the energies shown in Fig.~\ref{fig:2Fe2S} (b), the charge and spin observables are also in close agreement. The complete occupation numbers of all the atomic orbitals can be found in the Appendix.}

\begin{table*}[t!]
    \centering
    \begin{tabular}{l | c | c | c | c | c | c | c}
         \hline
         \hline
         \textbf{Method} &  \multicolumn{3}{|c|}{\textbf{Fe left}} & \multicolumn{3}{|c|}{\textbf{Fe right}} & \\
         \hline
          & N$_{e,\uparrow}$ & N$_{e,\downarrow}$ & Spin & N$_{e,\uparrow}$ & N$_{e,\downarrow}$ & Spin & Spin correlation \\
         \hline 
         \multicolumn{8}{l}{Restricted} \\
         \hline
         RHF            & 3.1075 & 3.1075 & 0        & 3.1071 & 3.1071  &  0          & -0.0437 \\
         MP2            & 3.1112 & 3.1112 & 0        & 3.1106 & 3.1106  &  0          & -0.0183 \\
         CISD           & 3.1337 & 3.1337 & 0        & 3.1327 & 3.1327  &  0          & -0.0560 \\
         CCSD           & 3.1397 & 3.1397 & 0        & 3.1393 & 3.1393  &  0          & -0.2085 \\
         \revision{HCI ($D=196.1$ M)} & 3.0103 & 3.0103 & 0        & 3.0089 & 3.0089  &  0          & -1.7807 \\
         \hline
         \multicolumn{8}{l}{Unrestricted} \\
         \hline
         UHF   & 4.9970 & 0.7993 & 2.0988 & 0.7966 & 4.9970 & -2.1001 & -0.0036 \\
         UMP2  & 4.9960 & 0.8285 & 2.0837 & 0.8258 & 4.9960 & -2.0851 & -0.0011 \\
         UCISD & 4.9734 & 1.0132 & 1.9801 & 1.0104 & 4.9735 & -1.9815 & -0.0112 \\
         UCCSD & 4.9515 & 1.0347 & 1.9583 & 1.0320 & 4.9516 & -1.9598 & -0.0288 \\
         \hline
         \multicolumn{8}{l}{Quantum} \\ 
         \hline
         SQD  ($D=16.7$ M) & 3.1121 & 3.1121   & 0 & 3.1116 & 3.1116 &  0   & -0.6315\\
         Ext-SQD ($D=126.2$ M) & 3.0130  & 3.0130 & 0 & 3.0118 & 3.0118 & 0 &  -1.7619 \\
         \hline
         \hline
    \end{tabular}
    \caption{\textbf{\revision{Ground-state charge and spin} properties of the [2Fe-2S] cluster.}
    Average number of spin-up and spin-down electrons and local spin-$z$ on the left (columns 2-4) and right (coulmns 5-7) Fe atom of the [2Fe-2S] cluster, and correlation between local spins on the two Fe atoms (column 8), from restricted (R, rows 4-\revision{8}) and unrestricted (U, rows \revision{10-3}) Hartree-Fock (HF), M{\o}ller-Plesset 2nd order perturbation theory (MP2), coupled-clusted singles and doubles (CCSD), and configuration interaction singles and doubles (CISD). 
    \revision{Row 8 reports the result for the largest HCI calculation shown in panel (a) of \cref{fig:2Fe2S}.}
    Rows 14 and 15 correspond to the quantities from SQD and Ext-SQD \revision{(SDT)}, respectively.}
    \label{tab:fe2s2_quantities}
\end{table*}

 \begin{table}[ht!]
    \centering
    \begin{tabular}{l | c | c | c | c | c | c | c}
         \hline
         \hline
         \textbf{Exc. } &  \multicolumn{3}{|c|}{\textbf{Fe left}} & \multicolumn{3}{|c|}{\textbf{Fe right}} & \\
         \hline
          & N$_{e,\uparrow}$ & N$_{e,\downarrow}$ & Spin & N$_{e,\uparrow}$ & N$_{e,\downarrow}$ & Spin & Spin corr. \\
         \hline 
         \multicolumn{8}{l}{\revision{HCI ($D=196.1$ M)}} \\
         \hline
         S$_1$       & 3.1030 & 3.1030 & 0        & 3.0767 & 3.0767  &  0      & -1.3600 \\
         S$_2$       & 3.0966 & 3.0966 & 0        & 3.1021 & 3.1021  &  0      & -1.1954 \\
         \hline
         \multicolumn{8}{l}{Ext-SQD ($D=126.2$ M)} \\
         \hline
         S$_1$       & 3.1219 & 3.1219 & 0        & 3.1216 & 3.1216  & 0       & -1.0102 \\
         S$_2$       & 3.0904 & 3.0904 & 0        & 3.0892 & 3.0892  & 0       & -1.3646 \\
         \hline
         \hline
    \end{tabular}
    \caption{\textbf{\revision{Excited-state charge and spin properties of excited states of the [2Fe-2S] cluster.}} \revision{Columns show the same physical observables as in Table~\ref{tab:fe2s2_quantities} for the S$_1$ and S$_2$ excited states both for HCI and Ext-SQD (SDT). }
    }
    \label{tab:fe2s2_exc_quantities}
\end{table}

\section{Conclusion and Outlook}
\label{sec:conclusion}

In this work, we explored the generalization of sample-based quantum diagonalization (SQD) to the calculation of molecular excited states. We considered three strategies: 
(i) the calculation of multiple Hamiltonian eigenpairs (beyond the ground-state one) within an SQD calculation, (ii) the quantum subspace expansion obtained applying single and double excitation operators (QSE(SD)), and (iii) the extension of SQD where single and double excitation operators are applied  to the individual configurations in the SQD wavefunction, yielding a larger set of configurations where used to compute multiple Hamiltonian eigenpairs (Ext-SQD).

Ext-SQD yielded the most accurate excited states, significantly outperforming SQD and QSE(SD) thanks to a greater variational flexibility. Finally, the computational overhead that Ext-SQD introduces over SQD is considerably modest compared to QSE(SD).

Because Ext-SQD returns approximate Hamiltonian eigenstates in the classically accessible form of sparse CI vectors (i.e. linear combinations of a polynomial number of configurations), users can easily certify and rank the quality of Ext-SQD excited states -- e.g., by computing their energy variances and mean energies~\cite{wu2023variational} -- and use Ext-SQD wavefunctions as input in further classical post-processing. This includes the computation of various excited-state properties, transition properties (e.g. 1-body reduced density matrices and dipoles), and frequency-dependent response functions (although it should be noted that a large number of excited states, often with energy well above that of T$_1$ or S$_1$, are necessary for accurate computation of these quantities).

The results of this study indicate that Ext-SQD can be used to compute singlet-singlet and singlet-triplet gaps relevant for, e.g., thermally activated delayed fluorescence~\cite{uoyama2012highly,gao2021applications} and photoisomerization~\cite{bylina1969photo,saltiel1975triplet} -- as well as other energy differences like ionization potentials, electron affinities, and band gaps.
In addition, due to the possibility of computing excited-state and transition properties, SQD can become a useful tool in the computation of frequency-dependent response functions -- including dipole spectra, density-density correlation functions, spin-spin 
correlation functions, and Green's functions. Importantly, Ext-SQD allows quantum computing studies in active spaces with considerably more electrons and orbitals than previously possible, because they employ classical and quantum computational resources in synergy and remove the need of measuring high-order density matrices, a bottleneck of widespread forms of quantum subspace calculations~\cite{motta2024subspace}.

Finally, we remark that Ext-SQD can be transported to Hamiltonians that are linear combinations of polynomially many Pauli or $k$-body fermionic operators (e.g. the Ising, Heisenberg, and Hubbard Hamiltonians) without major algorithmic modifications, and can be used to approximate their low-lying Hamiltonian eigenstates (assuming their sparsity). Therefore, it can be used beyond electronic structure and contribute more broadly to the simulation of many-body quantum systems on quantum computers.

\section*{Acknowledgements}

We acknowledge useful discussion with Antonio Mezzacapo and Kunal Sharma. The code used to produce the data in this manuscript is publicly available in the GitHub repository~\cite{sqd_addon}.

\clearpage
\newpage

\setcounter{equation}{0}
\setcounter{section}{0}  
\setcounter{table}{0}  
\setcounter{figure}{0}   
\renewcommand{\theequation}{A\arabic{equation}}
\renewcommand{\thefigure}{A\arabic{figure}}
\renewcommand{\thetable}{A\arabic{table}}

\appendix
\onecolumngrid
\section{Details of extended-SQD algorithm}
\label{app:tech_details}
This Appendix provides a detailed description of the extended-SQD procedure. 

\subsection{Extending the SQD subspace}

As explained in \cref{subsec:exc-state methods} the extended subspace of Ext-SQD is obtained by applying a set of excitation operators $\hat{E}_{I}$ to the configurations of the SQD wave function.
As we consider product of creation and annihilation operators, each excitation operator $\hat{E}_I$ on a computational basis state $\ket{\bts{y}_k}$ yields another computational basis state
\begin{equation}
\label{eq:dummy_eq}
\hat{E}_I \ket{\bts{y}_k} = \gamma_{Ik} \ket{\bts{z}_{Ik}}
\;,\;
\gamma_{Ik} \in \{0,1,-1\}
\;.
\end{equation}

The application of $\hat{E}_I$ to the configuration $\ket{\bts{y}_k}$ has four different outcomes:

\begin{enumerate}
    \item the configuration $\bts{z}_{Ik}$ is new to the subspace and unique
    \item the configuration $\bts{z}_{Ik}$ is not physical
    \item the configuration $\bts{z}_{Ik}$ is new to the subspace but equal to one already obtained $\bts{z}_{Jl}$
    \item the configuration $\bts{z}_{Ik}$ was already in the SQD subspace.
\end{enumerate}

While the first outcome is the one we are looking for and expected to provide the accuracy improvements we show in the main text, the other three have to be considered.
Given that we choose $\hat{E}_I$ to be particle-preserving, the outcome 2. is easily handled by checking that $\gamma_{Ik} = 0$.
Outcomes 3. and 4. can be controlled by checking the already existing configurations.
This check can be performed after every new configuration $\bts{z}_{Ik}$ is produced, which results in time consuming strictly non-parallel operations.
This check can also be executed after all the new configurations $\bts{z}_{Ik}$ are produced.
In this case, the prcedure is easily parallelizable, but produces a computational overhead on the memory used. 

In order to find a good tradeoff between the two cases, we apply the set of excitation operators on batches of configurations taken from the SQD wave function.
The size and the number of batches can be tuned depending on the system under study.
On a single batch, the application of the excitation operators is executed in parallel.
Then, the check is perfomed inter- and intra-batches serially.

Given a suitable amount of time, the application of excitation operators to produce the Ext-SQD can therefeore be performed on every classical device which is able to store the initial SQD wave function.
Finally, as indicated in \cref{subsec:exc-state methods}, the dimension of the extended subspace $D_E$ is is bounded by $D \prod_\sigma N_\sigma (M-N_\sigma)$.
Therefore, no asimptotically-hard procedure is introduced by the Ext-SQD workflow.

\subsection{Cutting the SQD subspace}

The SQD algorithm produces a classic vector representing the wave function of the physical system in the quantum-selected subspace of the Hilbert space.
As shown in \cite{robledomoreno2024}, SQD is able to produce wave function in very good qualitative agreement with the best classical solvers systems up to 77 qubits.
When we look at the accuracies of individual wave function amplitudes produced by SQD, the agreement is exceptional for the larger ones and slowly deteriorates in the description of the tails of the wave function.

While this is to be expected, we can use this behavior to speed up the Ext-SQD calculations without compromising the accuracy of the results.
Indeed, before extending the subspace with the excitation operators, one could choose to discard all the configurations that have amplitudes below a certain threshold, effectively reducing the size of the SQD subspace.

\begin{figure*}[ht]
    \includegraphics[width=1\textwidth]{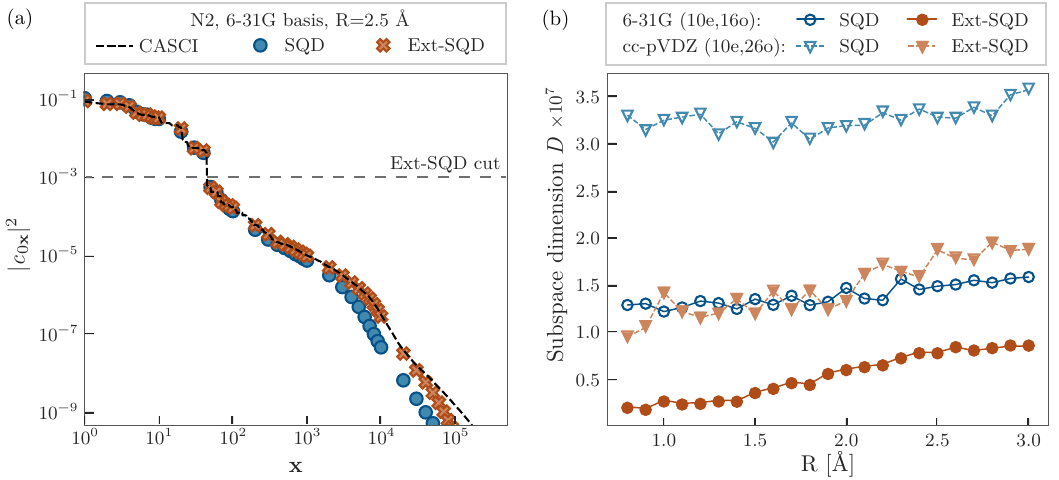} 
      \caption{
        \textbf{Cutting in N$_2$ dissociation.}
        \textbf{(a)} Shows for a selected configuration $R=2.5 \,\mathrm{\AA}$ the GS wave function amplitudes $|c_{0\textbf{x}}|^2$ for the corresponding electronic configuration $\textbf{x}$.
        SQD and Ext-SQD are compared to classical CASCI results.
        The grey dashed line indicates the threshold used to cut the SQD wave function before extending the subspace with single and double excitation operators.
        For better visualization clarity, the number of configurations shown has been reduced.
        \textbf{(b)} Shows the dimension of the subspaces obtained with SQD and Ext-SQD in the N$_2$ hardware experiments along the dissociation path.
      }
      \label{fig:cut_N2}
\end{figure*}

The choice of the threshold is dependent on the system under study.
For N$_2$ dissociation in cc-pVDZ (and in 6-31G, see \cref{app:sim_details}) basis set we always employed a threshold of $1 \cdot 10^{-3}$ on the wave function amplitudes for every configuration.
The loss in accuracy was always less than a mHa, a decrement well compensated by the final accuracy improvement of Ext-SQD.
Panel (a) of \cref{fig:cut_N2} compares the ground-state wave function amplitudes  $|c_{0\textbf{x}}|^2$ at $R=2.5 \,\mathrm{\AA}$ obtained with SQD and Ext-SQD, comparing them to CASCI results.
As shown in \cite{robledomoreno2024}, at large bond lengths, with strong static correlation effects, we expect the agreement between SQD and CASCI to be worse than at equilibrium.
We observe that below the threshold chosen the accuracy of SQD starts deteriorating, while Ext-SQD amplitudes follow CASCI ones down to $10^{-7}$.

Another benefit of the cut of the SQD wave function is illustrated in \cref{tab:sqd}: the final extended subspace can be smaller than the initial in certain instances.
As an example, the largest subspace obtained with Ext-SQD after cutting and extending is roughly a half of the largest SQD one, while the results are always an order of magnitude more accurate.
Panel (b) of \cref{fig:cut_N2} provides more details on the results obtained.
When the final subspace dimensions of SQD and Ext-SQD are compared, we notice that the final size of Ext-SQD calculations in cc-pVDZ basis set are similar to the size of SQD calculations in 6-31G basis, despite having 10 orbitals more.
These results highlight the capability of Ext-SQD to add to the final subspace configuration that are more relevant to ground and low-lying states calculations.

\section{Quantum experiment details}
\label{app:exp_details}
This appendix provides details about the experiments conducted on IBM's quantum devices. 

\begin{figure*}[ht]
    \includegraphics[width=1\textwidth]{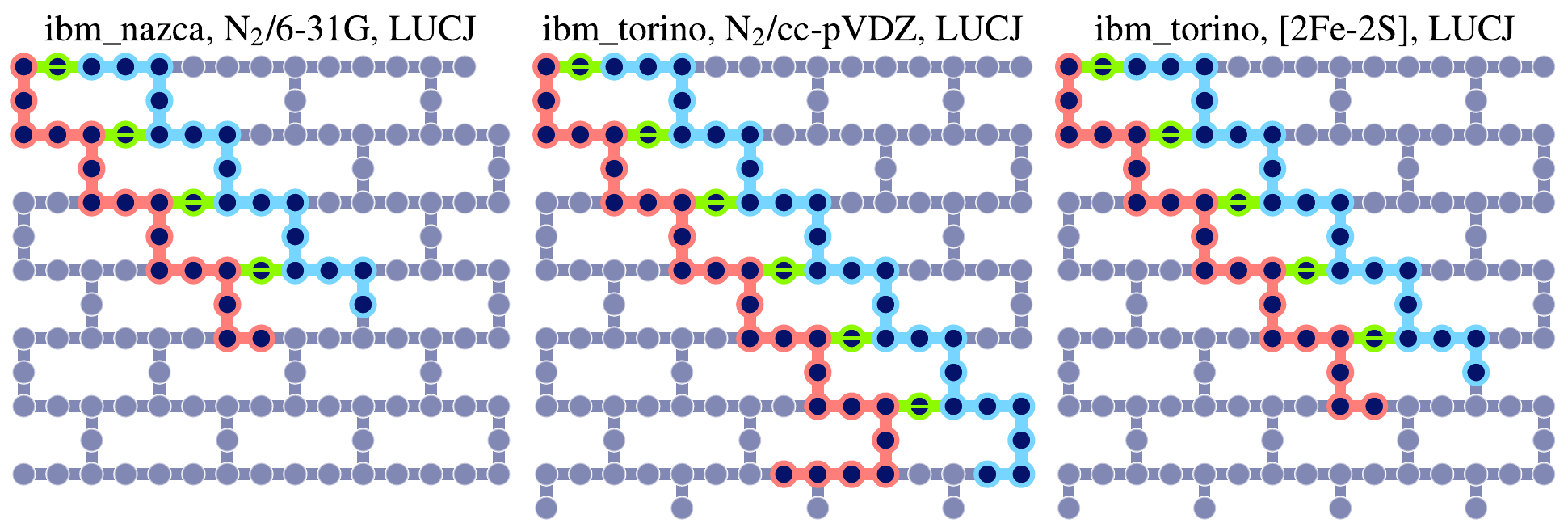} 
      \caption{
        \textbf{Experiment layout on IBM quantum processors.} Schematics of the devices used to carry out experiments for the N$_2$ molecule with 6-31G basis, the N$_2$ molecule with cc-pVDZ basis, and the [2Fe-2S] cluster. Qubits used in the calculation are shown in red (for qubits associated with $\alpha$ spin-orbitals), blue for qubits associated with $\beta$ spin-orbitals), and green (for auxiliary qubits).
      }
      \label{fig:device_layout}
\end{figure*}

The quantum circuits that were considered in these experiments are of the LUCJ type, as defined in Eq.~\ref{eq:lucj} in the main text. Their parameters are obtained from classical CCSD theory, following the procedure used in Ref.~\cite{robledomoreno2024}. Table~\ref{tab:hardware_details} provides a summary of the quantum resources involved in the experiments, specifically: the number of qubits and quantum gates and shots $|\tilde{\mathcal{X}}|$ for each circuit, the device used and the specific layout (see Fig.~\ref{fig:device_layout} for additional details), the fraction of of the configurations live in the subspace of the Fock space with the correct particle number, defined as:
\begin{equation}\label{eq:fraction_hardware}
    p_N^{\mathrm{hw}} = \frac{1}{|\tilde{\mathcal{X}}|} \sum_{\ell=1}^{|\tilde{\mathcal{X}}|} \delta_{N_{\bts{x}_\ell\alpha}, N_\alpha} 
\delta_{N_{\bts{x}_\ell\beta}, N_\beta},
\end{equation}
and the fraction of sampled configurations with the correct particle number if samples were collected from the uniform distribution over configurations of length $2M$
\begin{equation}\label{eq:fraction_uniform}
p_N^{\mathrm{unif}} = \binom{M}{N_\alpha} \binom{M}{N_\beta} 2^{-2M}.
\end{equation}
These two probabilities are statistically distinguishable for all the experiments we carried out.

Our experiments used twirled readout error mitigation (ROEM) \cite{nation2021scalable} to mitigate errors arising from qubit measurement. Dynamical decoupling (DD)~\cite{viola1998dynamical,kofman2001universal,biercuk2009optimized,niu2022effects} sequences of $X$ control pulses  to mitigate errors arising from quantum gates are also employed. We employed the implementation of ROEM and DD available on the \texttt{Runtime} library of Qiskit~\cite{Qiskit}, through the \texttt{SamplerV2} primitive.
The DD sequences consist  of two $X$ pulses (as in  Ramsey echo experiments) to idle qubits.

\begin{table*}
\begin{tabular}{cccccccc}
\hline
\hline
system & q (JW, Tot.) & ($d,\mathsf{CNOT},\mathsf{u})$ & device & layout & $p_N^{\mathrm{hw}}$ [95$\%$ c.i.] & $p_N^{\mathrm{unif}}$ & $|\tilde{\mathcal{X}}|$ \\
\hline
N$_2$, (10e,16o) & (32, 36) & (148,762,1408) & \device{nazca} & [0$\rightarrow$82]+[2$\rightarrow$73] & (0.0069,0.0071) & 0.0044 & \scientific{1}{5} \\
\hline
N$_2$, (10e,26o) & (52, 58) & (223,1792,3412) & \device{torino} & [0$\rightarrow$121]+[2$\rightarrow$127] & (0.0016,0.0017) & \scientific{9.6}{-7} & \scientific{9.8304}{4} \\
\hline
[2Fe-2S], (30e,20o) & (40, 45) & (173,1100,2070) & \device{torino} & [0$\rightarrow$104]+[2$\rightarrow$94] & (0.0044,0.0045) & 0.00022 & \scientific{2.4576}{6} \\
\hline
\hline
\end{tabular}
\caption{Hardware details. For the active space of each molecular system (first column), we list the details of the LUCJ quantum circuit executed on hardware: its number of qubits (q) both for the Jordan-Wigner encoding (JW) and the total number of qubits used including auxiliary ones (Tot.) (column 2) and count of quantum operations (circuit depth $d$, number of $\mathsf{CNOT}$ and single-qubit $\mathsf{u}$ gates, column 3), the device used (column 4) and the qubit layout chosen (column 5, where $[a\rightarrow b]$ indicates the shortest path between qubits $a$ and $b$ in the device topology (see Fig.~\ref{fig:device_layout}). In columns 6 and 7 we report a 95\% confidence interval for the fraction $p_N^{\mathrm{hw}}$ of configurations with the correct particle number and the corresponding value for configurations  with uniform probability distribution (see Eqs.~\ref{eq:fraction_hardware} and~\ref{eq:fraction_uniform}). In column 8, we list the number of raw measurement outcomes obtained from the device.}
\label{tab:hardware_details}
\end{table*}

\section{Additional results}
\label{app:sim_details}
This Appendix provides extra experimental results.

\subsection{Dissociation of nitrogen using 6-31G}

\begin{figure*}[ht]
    \includegraphics[width=1\textwidth]{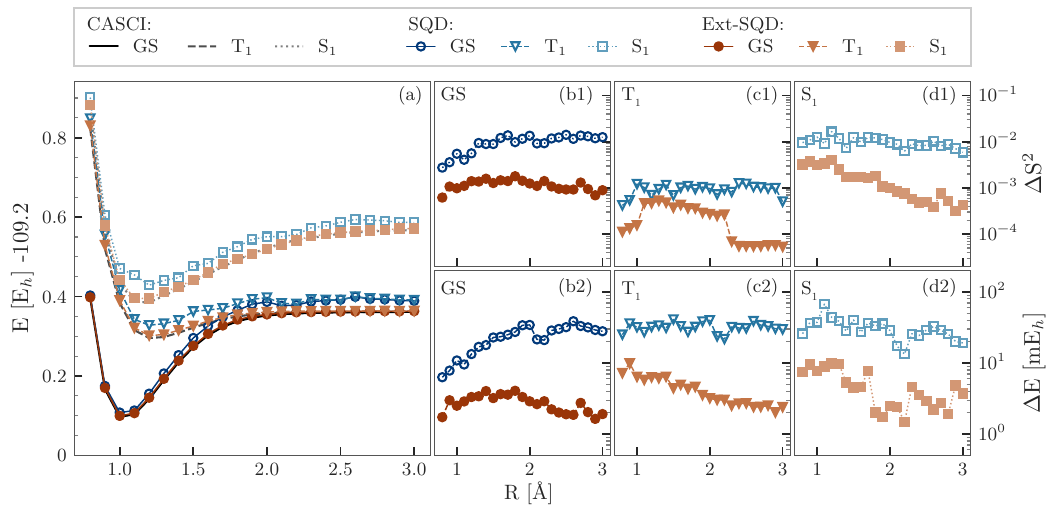} 
      \caption{
        \textbf{N$_2$ dissociation.}
        \textbf{(a)} Ground-state (GS) and first two excited-state potential energy curves (T$_1$ and S$_1$) of N$_2$ in the 6-31G basis set from SQD, extended SQD (Ext-SQD), and heat-bath configuration interaction (HCI).
      \textbf{(b1, c1, d1)} Deviations $\Delta S^{2}$ from HCI in the total spin, for SQD and Ext-SQD, along the GS (b1), T$_1$ (c1), and $S_1$ (d1) potential energy curves.
      \textbf{(b2, c2, d2)} Deviations $\Delta E$ from HCI in the total energy, for for SQD and Ext-SQD, along the GS (b2), T$_1$ (c2), and $S_1$ (d2) potential energy curves.
      }
      \label{fig:diss_N2_631g}
\end{figure*}

Hardware experiments for N$_2$ dissociation were performed in the 6-31G basis set with frozen core approximation.
With a total of 10 electrons in 16 orbitals, these experiments represent a intermediate step between the (10e, 8o) and (10e, 26o) active spaces of cc-pVDZ  basis set.
We collected $|\tilde{\mathcal{X}}|=10^5$ samples from \device{nazca}, using LUCJ circuits and 36 qubits.
The final SQD and Ext-SQD results are shown in \cref{fig:diss_N2_631g}, where they are compared against exact classical calculations performed in the same basis set (CASCI).

\subsection{\revision{Orbital occupancy profiles for [2Fe-2S] cluster}}

\begin{figure*}[ht]
    \includegraphics[width=1\textwidth]{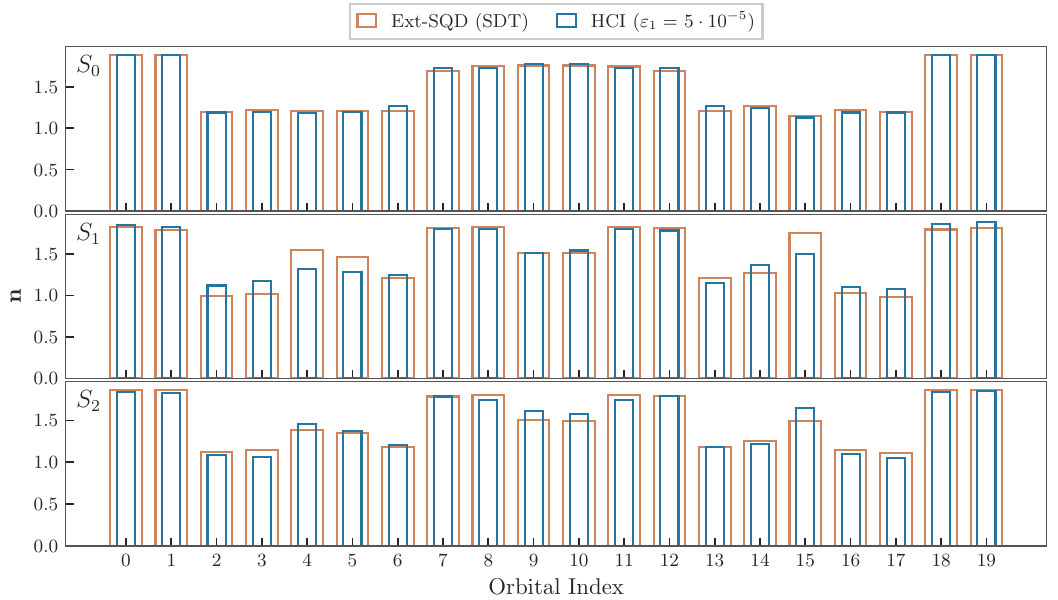} 
      \caption{
        \textbf{[2Fe-2S] occupancy profiles.}
        \revision{Orbital occupancy profiles $\mathbf{n}$ for the ground state ($S_0$) and the first two singlet excited states ($S_1$ and $S_2$) of the [2Fe-2S] cluster. 
        The occupancies obtained with Ext-SQD with the inclusions of singles, doubles, and triples are compared to the Heat-Bath CI results (see main text for more info on the methods).}
        }
      \label{fig:fes_occupancy}
\end{figure*}

\revision{In the main text we reported the average number of Fe electrons for the ground state of the [2Fe-2S] cluster and its two excited states $S_1$ and $S_2$.
Here, for completeness, we report in \cref{fig:fes_occupancy} the full occupancy profile $\mathbf{n}$ for each of these states. 
A representation of the orbitals can be found in \cite{robledomoreno2024,li2017spin}.
}

\end{document}